# High-Throughput Transaction Executions on Graphics Processors

Bingsheng He
Nanyang Technological University, Singapore

Jeffrey Xu Yu
The Chinese University of Hong Kong

## ABSTRACT

OLTP (On-Line Transaction Processing) is an important business system sector in various traditional and emerging online services. Due to the increasing number of users, OLTP systems require high throughput for executing tens of thousands of transactions in a short time period. Encouraged by the recent success of GPGPU (General-Purpose computation on Graphics Processors), we propose *GPUTx*, an OLTP engine performing high-throughput transaction executions on the GPU for in-memory databases. Compared with existing GPGPU studies usually optimizing a single task, transaction executions require handling many small tasks concurrently. Specifically, we propose the *bulk execution model* to group multiple transactions into a bulk and to execute the bulk on the GPU as a single task. The transactions within the bulk are executed concurrently on the GPU. We study three basic execution strategies (one with locks and the other two lock-free), and optimize them with the GPU features including the hardware support of atomic operations, the massive thread parallelism and the SPMD (Single Program Multiple Data) execution. We evaluate GPUTx on a recent NVIDIA GPU in comparison with its counterpart on a quad-core CPU. Our experimental results show that optimizations on GPUTx significantly improve the throughput, and the optimized GPUTx achieves 4–10 times higher throughput than its CPU-based counterpart on public transaction processing benchmarks.

## 1. INTRODUCTION

OLTP (On-Line Transaction Processing) is an important business sector generating billions of dollars revenues for database vendors, and even more for online service providers. The market for OLTP is ever growing, as the volume of traditional applications such as credit card services, banking and stock markets becomes larger, and emerging applications including Web 2.0 become popular. In those applications, tens of thousands of transactions are required to be processed within a short period. Additionally, transactions are usually implemented as stored procedures without any user interaction stalls [18]. In those applications, system optimizations are throughput oriented, instead of response time oriented. This calls for a high-throughput transaction execution engine. Most current CPU-based systems [18, 7] adopt in-memory solutions, and utilize multiple CPUs or multiple commodity machines to achieve the performance requirement. Encouraged by the recent success of GPGPU (General-Purpose computation on Graphics Processors), we investigate whether and how we can design a transaction execution engine on the GPU for high throughput.

GPUs (Graphics Processors) have evolved as many-core processors for general purpose computation. They have over an order of magnitude higher memory bandwidth and higher computation power (in terms of GFLOPS) than CPUs. The superb hardware resource enables a GPU to concurrently execute tens of thousands of threads, which can effectively hide the memory latency [8]. This massive thread parallelism is an ideal hardware advantage for transaction executions, where data accesses are usually random and the overall performance is memory latency bounded. While current GPGPU techniques accelerate various database tasks [8, 12], they are mostly limited to a single task such as sort [4] and joins [10], which perform read-only computation on a large volume of data. In comparison, OLTP systems need to handle many small transactions with random reads and updates on the database. Moreover, transaction executions must achieve isolation and consistency for correctness when they perform concurrent updates to the database.

Unique features of transaction processing distinguish itself from the existing GPGPU research. The massive thread parallelism of the GPU poses technical challenges on the correctness and efficiency of transaction executions. On the GPU, executions with thousands of threads are organized in the SPMD (Single Program Multiple Data) execution model. The SPMD execution model has important implications to the efficiency of transaction executions. First, ad-hoc transaction execution (i.e., handling one transaction at a time) is not desirable, since it causes severe underutilization of the GPU. Even with the support of concurrent task (or *kernel*) executions on the future GPU, ad-hoc transaction execution still causes underutilization, since each transaction occupies at least one multi-processor. Second, the atomic operations supported in new generation GPUs enable us to implement locks for concurrency control. However, the performance of locking mechanisms should be revisited under the context of the SPMD execution and the massive thread parallelism of the GPU.

To address these challenges, we propose the *bulk execution model* for transaction executions on the GPU. In the model, a transaction belongs to a specific transaction type. Each transaction type is registered as a stored procedure without user interaction, and the codes of registered transaction types are combined into a single kernel. A transaction is an instance of the registered transaction type with different parameter values. Thus, a set of transactions can be grouped into a single task, namely a *bulk*. Transactions within a bulk are concurrently executed on the GPU.

Such a simple execution model enables various optimization opportunities for addressing the technical challenges. First, the bulk execution model allows much more concurrent transactions than ad-hoc execution. The elimination of user interaction latency tends to keep the processor busy. Second, data dependencies and branch divergence among transactions are explicitly exposed within a bulk. With known data dependencies, the system can choose the suitable concurrency control strategy to execute the bulk, e.g., with and without locks. Considering branch divergence, transactions with the same type can be grouped together for execution, in order to minimize the overhead of branch divergence. Third, transaction executions become tractable. Instead of executing ad-hoc transactions, concurrent executions occur within a single kernel.

To capture the data dependency for the transactions in the bulk, we propose a data structure called *T-dependency graph*, where we augment the classic dependency graph with the timing of the transaction submissions. The T-dependency graph is a guide for the bulk execution model, ensuring that the result of transaction executions in the bulk is equivalent to that of sequentially executed the transactions in the bulk according to their submission time. On the other hand, T-dependency graph formally exposes the parallelism within the bulk. For example, we can identify the set of transactions that do not have any preceding conflicting transactions. These transactions can be executed in parallel without any complicated concurrency control.

Based on the bulk execution model and T-dependency graph, we develop *GPUTx*, a transaction processing engine prototype on the GPU for in-memory databases. We study three strategies for bulk execution, one with traditional two phase locking mechanism, or two lock-free algorithms. One lock-free algorithm adopts the partition-based transaction execution [18], and the other always executing the set of transactions that do not have preceding conflicting transactions according to the T-dependency graph. All these execution strategies are optimized with the massive thread parallelism and minimizing the branch divergence.

We conduct experiments on a NVIDIA GPU of 240 cores with synthetic workloads and public OLTP benchmarks such as TPC-C and TM1 [13]. We compare the performance of GPUTx with our homegrown CPU-based counterpart. The results show that: a) the optimization techniques improve the transaction execution throughput, e.g., minimizing the branch divergence can achieve up to an order of magnitude throughput improvement; b) the bulk execution model improves the throughput of GPU-based transaction execution, 16–146 times higher than ad-hoc transaction execution on the GPU; c) GPUTx achieves a 4–10 times higher throughput than its CPU-based counterparts on the quad-core CPU.

**Organization.** The remainder of this paper is organized as follows. Section 2 reviews the related work on the CPU-based transaction processing and the GPU-based query processing. Section 3 gives an overview of the system design. We present the T-dependency graph in Section 4, followed by the implementation details in Section 5. We present the evaluation results in Section 6, and discuss the limitations of GPUTx in Section 7. Finally, we conclude in Section 8.

## 2. PRELIMINARY AND RELATED WORK

**Transaction Processing on CPUs.** Main memory databases [7, 9] have recently attracted a significant amount of attention due to the great increase of the memory capacities. In-memory transaction engines such as H-Store [18] and StagedDB [7] has become popular in recent years.

Efficient transaction executions on multi-core CPUs is a hot research topic [18, 11, 16]. Harizopoulos et al. [6] performed a detailed performance study on the components in a memory-resident transaction processing system, and showed that only a small portion of the total time is spent on the useful work. H-Store [18] models a transaction as a stored procedure. It uses data partitioning and single-threaded execution to simplify concurrency control (e.g., removing two-phase locking mechanism [17]). Jones et al. [11] studied the concurrency control performance on a partitioned database. Along the line of improving the efficiency of transaction executions, DORA [16] is a data-oriented execution engine, re-designed for multi-core CPUs. In DORA, transaction executions are aligned to their target data items. This can reduces the ad-hoc locking operations. All of these studies adopt ad-hoc execution models on multi-core CPUs. Due to the differences in hardware architecture and execution model, these results need to be revisited, and new techniques should be proposed for bulk transaction executions on the GPU.

**Query processing on the GPU.** GPGPU has been a fruitful research area in recent years. We refer the readers to Appendix A for the preliminary on graphics processors, and a survey [15] for the details on GPGPU applications.

The superb raw hardware power of the GPU has been exploited to accelerate various applications in databases [5, 10, 8, 2] and other data-intensive tasks [1]. All of these studies focus on OLAP-like applications, each task usually processing a large volume of data. In comparison, this study focuses on another important area in databases, OLTP, which receives little attention in GPGPU. In an OLTP application, we are facing tens of thousands of small tasks in a short time, and a high-throughput engine becomes a necessity. One of the possible reasons that OLTP receives little attention in GPGPU is that OLTP is considered to be ad hoc due to the latency in user interaction. However, in practice, many OLTP applications (such as Web 2.0) become throughput oriented, without user interaction latency [18]. GPUTx is specially designed and optimized for such applications.

The recent hardware support for atomic operations on the GPU enables us to explore concurrent transaction executions with locks. To the best of our knowledge, there is not yet any transaction processing engine on the GPU.

## 3. SYSTEM DESIGN

### 3.1 Bulk Execution Model

In the bulk execution model, a bulk is defined to be a set of transactions. Bulk execution model only supports

315

pre-defined transaction types. Each transaction type is registered as a stored procedure without user interaction. A transaction is an instance of a certain transaction type with parameter values. That is, GPUTx supports the predefined transaction:

`Execute procedure_name (parameter list)`

Each transaction is associated with a timestamp indicating the time when it is submitted to the system.

Bulk execution is to execute the transactions within the bulk according to a certain execution strategy. In addition to the isolation and consistency requirement on individual transactions, the bulk execution has its own requirement for correctness. Specifically, we define the correctness of bulk execution in Definition 1. While the timing constraint is not necessary for serializability of transaction executions, it ensures the correctness of executing a large bulk.

**Definition 1.** *Given any initial database, a bulk execution is correct if and only if the result database is the same as that of sequentially executing the transactions in the bulk in the increasing order of their timestamps.*

### 3.2 Overview of GPUTx

GPUTx is a transaction processing engine, specially designed and optimized for the bulk execution model the GPU. Our current focus is on the data when they can fit into the GPU device memory.

Current GPUTx is implemented in NVIDIA CUDA [14]. A transaction type is implemented as a *device* function in CUDA. All the registered transaction types form a kernel function, connecting the stored procedure of each transaction type with a *switch* clause as followed. $type\_i$ is the type identifier for the $i$th transaction type.

```
switch(type) {
   case type_1:
       //stored procedure for type_1;
   break;
   ...
   case type_N:
       //stored procedure for type_N;
   break;
}
```

If a new transaction type needs to be registered to GPUTx, we add the stored procedure into the switch clause, and recompile the kernel function.

When transactions are submitted by users (e.g., via internet), their signatures are temporarily stored in a transaction pool. The transaction signature is represented in the form of <*id, type, parameter value list*>, where *id* is an unique and auto-increment identifier for each transaction, and *type* is the transaction type. We use the transaction ID to represent its timestamp.

The system periodically generates a bulk by picking a set of transactions from the transaction pool, and issues the bulk to the GPU for execution. When the execution is done, the results are copied from the GPU to the result pool in the main memory, and then returned to individual users.

GPUTx use arrays in the device memory to store the relation. For transactions with insertions, we allocate a temporary buffer that is sufficiently large for the new inserted data. After the kernel execution, we perform a batched update with updates in the temporary buffer.

Data accesses in GPUTx are performed at the granularity of data field, in order to maximize the parallelism on the GPU. This is in contrast with the traditional CPU-based approaches [16, 18], where the granularity is usually page- or partition-level.

## 4. T-DEPENDENCY GRAPH

### 4.1 Definitions

A transaction consists of multiple basic operations and their processing on databases. We define a basic operation to be a read or a write on a data item in databases. Since we assume there is no user interaction latency within the transaction, we assume that basic operations have the same timestamp as the transaction. We refer two basic operations are conflicting if they target at the same data, and at least one of them is write. We further define that two transactions $t_1$ and $t_2$ are conflicting if there are two conflicting basic operations $o_1$ and $o_2$ in transactions $t_1$ and $t_2$, respectively.

A correct bulk execution must take into account the conflicting transactions. Suppose two basic operations, $o_1$ and $o_2$ from two transactions are conflicting. If $o_1$ has a timestamp smaller than $o_2$, a correct bulk execution should guarantee that $o_1$ is performed before $o_2$. This property is to ensure the correctness of bulk execution, where the result of the bulk execution is the same as the sequential execution for the transactions in the bulk. Additionally, while this property confines the execution order of the conflicting operations, it indicates the opportunity of parallelism for identifying the transactions without any preceding conflicting transactions. These transactions can be executed in parallel without any complicated concurrency control.

We propose a data structure named *T-dependency graph* to explicitly capture the data dependency and correctness of bulk execution. A T-dependency graph is a DAG (Directed Acyclic Graph). Each vertex represents a transaction, and an edge represents the data dependency between two transactions. An edge is added to two vertexes ($t_1 \rightarrow t_2$) if and only if the following three conditions are all satisfied: (a) $t_1$ and $t_2$ are conflicting transactions, (b) $t_1$ has a smaller timestamp than $t_2$, and (c) there does not exist any transaction $t$ with a timestamp between those of $t_1$ and $t_2$ such that $t$ is conflicting with both $t_1$ and $t_2$. Condition (a) is to express the data dependency between two transactions. Condition (b) is according to the correctness of bulk execution. It ensures that there is no cycle in the T-dependency graph. Condition (c) ensures that an edge is added to two intermediate conflicting transactions. Figure 1 (a) illustrates the T-dependency graph for four transactions, T1, T2, T3 and T4 (in the order of increasing timestamps). While T1 and T4 are conflicting transactions, there is no edge between them, due to the violation on Condition (c).

After representing all the transactions in the transaction pool as a T-dependency graph, the data dependency and the timing relationship are explicitly exposed. For example, the set of vertexes without any preceding vertexes in the T-dependency graph indicates that their corresponding transactions do not have any preceding conflicting transactions. We define this kind of vertexes to be *source*. Based on sources, we define the *depth* of a vertex $v$ to be the length of the longest path from a source vertex to $v$. A source vertex has a depth of zero. We define the depth of the T-dependency graph to be the maximum depth of all vertexes in the graph. We further define $k$-set ($k \geq 0$) to be the set of vertexes with depth of $k$. Thus, all the source vertexes form the 0-set. In the example T-dependency graph of Figure 1, the 0-set contains T1, the 1-set includes T2 and T3, and the 2-set contains T4.

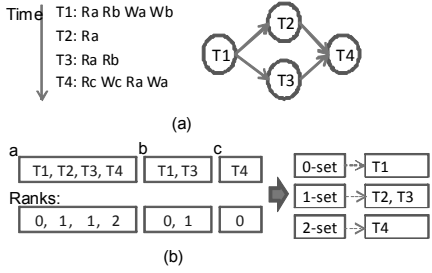

Figure 1: The T-dependency graph. R$x$ and W$x$ represent a read and a write on $x$, respectively.

By definition, we have the following two properties for the $k$-sets. Property 1 indicates that transactions from the same $k$-set are conflict-free, and a bulk consists of the $k$-set only can be executed in parallel while preserving the correctness. Property 2 indicates that we cannot execute the entire $k$-set as a bulk, if there are undone transactions in 0-set, 1-set, ..., $(k-1)$-set.

**Property 1.** $\forall t_k \in k\text{-set}$ and $\forall t'_k \in k\text{-set}$ ($k \geq 0$, $t_k \neq t'_k$), $t_k$ and $t'_k$ are conflict-free.

**Property 2.** $\forall t_k \in k\text{-set}$ ($k \geq 1$), there exists at least one $t_{k-1}$ in $(k-1)\text{-set}$ such that $t_{k-1}$ and $t_k$ are conflicting transactions.

The T-dependency graph is similar to the traditional dependency graph in DBMSs [17], since both of them aim at representing the data dependency among transactions. Our definition of the T-dependency graph is specific to the bulk properties. We consider the timestamp of a transaction in the T-dependency graph. Moreover, there is no cycle in the T-dependency graph, thus avoiding the deadlock in execution, whereas the traditional dependency graph may have cycles.

### 4.2 Operations

We introduce some basic operations on the T-dependency graph, including constructing T-dependency graph and calculating $k$-set. The details of the T-dependency graph construction can be found in Appendix B. We focus on the $k$-set calculation.

If the T-dependency graph has already been constructed, the $k$-set can be obtained by performing a topological sort on the T-dependency graph. During the topological sort, we evaluate the depth of a vertex $n$ to be $(1+d)$, where $d$ is the maximum depth of all the preceding vertexes of $n$. However, the $k$-set calculation is a basic building block for bulk execution (see Section 5). The T-dependency graph construction can become the bottleneck of the system. Thus, we develop an efficient algorithm for calculating the $k$-set on the GPU, without constructing the T-dependency graph.

Our algorithm is data oriented. We consider all the basic operations in the transaction pool. We first group the basic operations according to their target data item. After grouping, each group contains the potentially conflicting basic operations. Next, within each group, we further order the basic operations in the ascending order of their timestamps, and analyze the data dependency within the group. We assign a rank value for each basic operation in each group. The first basic operation in each group is assigned with a rank zero. For the $i$th basic operation ($i \geq 1$), we will look for the conflicting operation with a smaller ID in the group. We denote $r$ to be the rank value of the $(i-1)$ transaction. If the current basic operation is a write, the rank value of the $i$th operation is $(r+1)$. Otherwise, if $(i-1)$th operation is a read, we assign the rank value of the $i$th operation to be $r$. Otherwise, we assign the rank value of the $i$th operation to be $(r+1)$. We calculate the $k$-set according to the rank values of the basic operations in the transaction. A transaction belongs to $k$-set if and only if the maximum rank value of its basic operations is $k$.

Figure 1(b) demonstrates the process of calculating the $k$-set for the four example transactions. First, transactions form three groups according to their accesses to $a$, $b$ and $c$. Next, the ranks are given to the operations within the group. Take the group for $a$ as an example. Initially, $T_1$ has a rank of zero. Since $T_1$ has a write and $T_2$ has reads only, the rank for $T_2$ should be one plus the rank of $T_1$, i.e., one. For $T_3$, since both $T_2$ and $T_3$ are read-only, they have the same ranks. Thus, $T_3$ has a rank of one. Finally, $T_4$ has a write on $a$, and its rank should be one plus the rank of $T_3$, i.e., two. With all the ranks in the three groups, we know $T_1$ belong to 0-set, $T_2$ and $T_3$ belong to 1-set and $T_4$ belong to 2-set.

We represent a basic operation in a transaction to be a tuple $(v, id)$, where $v$ denotes an ID of the target data item, and $id$ is the transaction ID. Taken an array of such tuples as input, the GPU-based implementation works in the following five steps.

1) We use a sort operation to perform grouping on the array, firstly on $v$ and then on $id$.
2) We use a map primitive to identify the boundary of the groups.
3) We use a thread to evaluate the rank within each group. This step outputs the result of an array with element ($id$, $r$), where $id$ represents the transaction ID and $r$ denotes the rank value of each operation.
4) A sort operation is performed on the output array of Step 3).
5) We use a map primitive to identify the boundary of the groups. The ending element of each boundary represents the maximum rank of a transaction, i.e., the depth of the transaction in the T-dependency graph.

The implementation reuses existing efficient data-parallel primitives on the GPU, and exploits the parallelism among different groups.

## 5. IMPLEMENTATIONS

GPUTx takes transactions in the transaction pool as input, generates the bulk, performs the execution on the GPU, and outputs the results to the result pool. A bulk profiler is to analyze the characteristics of the input transactions. A bulk generator is to decide the suitable execution strategy, and to generate a bulk for execution. Taking a bulk as input, a bulk executor executes the bulk on the GPU, and returns the results.

The current bulk executor of GPUTx supports the following three basic execution strategies on a bulk.
• TPL: We adopt the classic two phase locking execution method, where locks are implemented with atomic operations on the GPU.
• PART: We adopt the partitioned based approach in H-Store [18], and a single thread is used for each partition.

Since it is single-thread execution, there is no locking mechanism within a partition.

• K-SET: K-SET is specially designed for bulk execution. It is based on the concept of $k$-set of the T-dependency graph, and identifies the transactions that do not have any preceding conflicting transactions to execute in parallel.

All these three strategies achieve the correctness requirements of bulk execution. In Appendix G, we briefly discuss the execution strategies of achieving the serializability only, i.e., relaxing the timestamp constraint on bulk execution.

## 5.1 TPL: Two phase locking

Two phase locking is a basic and widely used concurrency control method. According to the locking protocol, a transaction handles its locks in two phases: in the first phase, locks are acquired and no locks are released; and in the second phase, locks are released and no locks are acquired.

The naive method uses the GPU-based atomic operation to implement a 0/1 spin lock. The details can be found in Appendix C. With the spin lock, we can implement a two phase locking algorithm for transaction executions. While this implementation is simple, it inherits the obstacles of two phase locking, i.e., deadlocks and non-determinism in the execution order. On the GPU, we do not have explicit control for thread scheduling. Thread executions are non-deterministic with the simple 0/1 spin lock. For example, a transaction with a larger timestamp may be executed earlier than its conflicting transactions. Moreover, a deadlock causes a never terminated kernel execution.

We address these two obstacles based on the T-dependency graph. The non-determinism is caused by the mismatch between the execution order and the timestamp. To ensure the execution order, we extend the spin lock with multiple counter values. Before a basic operation accesses the shared data, it is assigned a key value for the lock. Only when the key value is equal to the counter value, the thread can acquire the lock. Thus, the key values are used to generate an order among thread executions. In order to achieve the correctness of the bulk execution, the key values are assigned as the rank of each group in the third step of the $k$-set calculation (Section 4.2). Through assigning the key values according to the rank value, the deadlock is avoided. This is because, T-dependency graph is a DAG without cycles. The detailed code lines are shown in Appendix C.

In the basic TPL, we need a spin lock for every data access. This strict requirement has overhead in assigning the key values to transactions, as well as runtime overhead in the spin lock. We should eliminate unnecessary locks whenever practical. For example, transactions in OLTP applications tend to fetch a small number of tuples according to the primary key. Moreover, schemas of many OLTP applications are tree shaped. The transactions, such as those in public benchmarks, belong to this category. The primary key of the root relation in the tree is used as the object for locking. A running example for TPC-B is shows in Figure 2(a), where there are $n$ bank branches, $B_1, B_2, ..., B_n$. Since transactions update the balance of a branch, any two transactions for the same branch are conflicting. In this case, the T-dependency graph degrades to be multiple paths corresponding to $B_1, B_2, ..., B_n$.

If we put a spin lock on the primary key value, we can eliminate locks to conflicting operations on the data items other than primary key. The script numbers in Figure 2(b) illustrate the key values assigned to the transactions in a bulk. The key values for the transactions in the $k$-set are $k$.

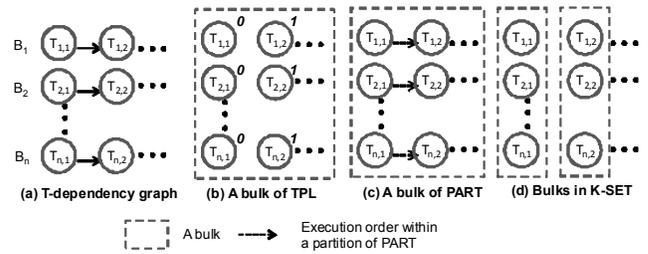

Figure 2: A running example for transaction execution strategies using TPC-B.

## 5.2 PART: Partition based execution

On the partitioned database, a single thread is used to handle the data operations to each partition. Since these operations are done sequentially on the target partition, there is no locking mechanism for the execution. This is particularly ideal for single-partitioned transactions (accessing the data entirely within a single partition), without complicated multi-partition concurrency controls or locks.

The parallelism is achieved by the concurrent executions on multiple partitions. In contrast with the CPU-based engines (like H-store) that assign the transactions to the suitable worker threads (i.e., a push model), the GPU-based execution is a pull model, where each GPU thread needs to identify the set of transactions belonging to itself.

The basic process of PART is implemented as the following three steps. First, we use a map primitive to calculate the partition ID of each transaction. The results are stored in the array $P$, where $P[i]$ is a tuple with the partition ID and the pointer in the input transaction fingerprint. Second, we sort the $P$ according to the partition ID with radix sort. Third, each GPU thread picks the transactions from $P$. It identifies the start and the end boundaries for its partition in $P$ with binary searches. Given the start and the end positions, the GPU thread sequentially executes the transactions in the partition. Figure 2(c) shows an example of PART. Each thread is responsible for a path in the T-dependency graph, corresponding to a partition.

The partition size is a tuning parameter for the performance of PART. If we increase the partition size, the number of partitions is reduced, and thus the sorting algorithm in the second step is more efficient. Moreover, the cost of the third step of picking the boundary can be reduced. However, a large partition size increases the number of transactions processed by a thread, which increases the length of the critical path of the entire kernel. We tune the suitable partition size for the optimal performance in the experiment.

PART works correctly on single-partitioned transactions. If there are cross-partition transactions, we use TPL for execution, which can severely degrade the performance. A more advanced concurrency control on cross-partition transactions should be investigated in the future.

## 5.3 K-SET: $k$-set based execution

The K-SET algorithm is based on the $k$-set concept of the T-dependency graph. It iteratively pick the 0-set as a bulk for execution. Since transactions in the 0-set do not have conflicts with each other, we do not need to use the mechanisms in TPL or PART. All the transactions in the bulk are executed in parallel on the GPU. If the bulk size is sufficiently large, the GPU computation resource can be fully utilized.

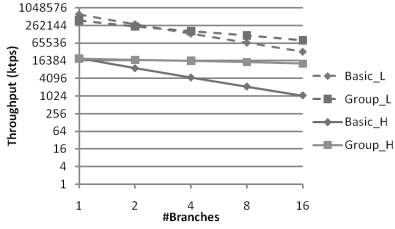

Figure 3: Throughput comparison varying the number of branches in the *switch* clause.

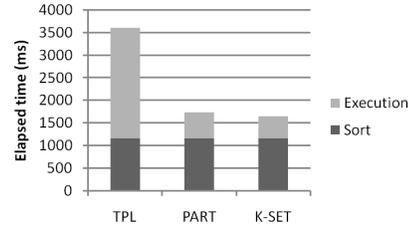

Figure 5: The time breakdown of the three transaction execution strategies.

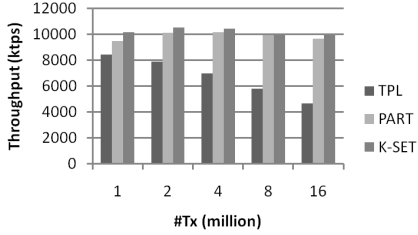

Figure 4: Throughput of the three transaction execution strategies varying the bulk size.

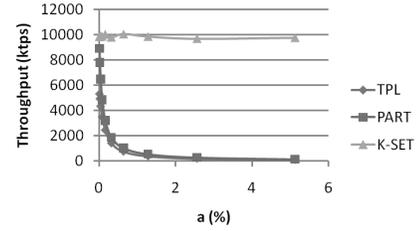

Figure 6: Throughput of the three transaction execution strategies varying the workload distribution.

The $k$-set calculation has been presented in Section 4.2. Incremental algorithms for the $k$-set calculation is adopted. When new transactions are added to the pool, their basic operations are merged into the sorted array. Next, we can select the bulk for the transactions with the key value of zero. The incremental algorithm is able to find the 0-set without computing the $k$-set from scratch.

Figure 2 (d) shows an example of running K-SET. We pick 0-set as a bulk. After removing the transactions from 0-set, the transactions in 1-set become the 0-set in the new T-dependency graph.

### 5.4 Optimization Issues

GPUTx embraces a number of optimization issues. In addition to the general optimization techniques (such as those on the memory locality and the thread parallelism [8]), we discuss the techniques that are mostly relevant to the bulk execution model and transaction processing. The details of these optimization issues can be found in Appendix D.

• In order to reduce the branch divergence, we group the transactions in a bulk according to their transaction types. The number of groups is tuned for the tradeoff between the grouping overhead and the gain of reduced branch divergence.

• Choosing the suitable execution strategy is important for the system throughput. We analyze the strength and weakness of the three execution strategies and use a rule-based method to determine the suitable execution strategy.

• Logging as well as the recovery overhead are eliminated in the system whenever practical.

## 6. EVALUATIONS

### 6.1 Experimental Setup

We run our experiments on a machine with four NVIDIA C1060 GPUs and one Intel Xeon CPU E5520. While the machine has four GPUs, they are independent in the system and we use only one of them for evaluating the performance of GPUTx. A detailed experimental setup is found in Appendix E.

Our experiments include two parts. The first part is to develop micro benchmarks to conduct controlled studies on the key design of GPUTx. The second part is to provide an end-to-end comparison of GPUTx to its CPU-based counterpart with three benchmarks namely TM1 [13], TPC-B, and TPC-C. These benchmarks represent different characteristics to assess the optimization opportunities of GPUTx. We mainly measure the long-running system throughput (ktps, or thousands of transactions per second).

In the micro benchmarks, we have varied the number of branches in a *switch* clause, $T$, and varied the amount of computation in each branch to evaluate branch divergence. Transactions are evenly assigned with a transaction type. We have examined the complied code, and made sure the branches are not eliminated by code optimization. Each transaction reads a tuple, and performs computation, and then writes the result back to the tuple. The amount of computation is simulated with calling the $\_\_sinf$ function ($100 \cdot x$) times. The default values for $T$ and $x$ are eight and 16, respectively. We further generated a skewed distribution on the lock acquisition among transactions. The skewedness is according to the $\alpha$ value, where transactions acquire the first lock with a probability of $\alpha$, and the probabilities of acquiring other locks are the same. A larger $\alpha$ value means a more skewed distribution, and thus a deeper T-dependency graph.

We present the key results on GPUTx. More experimental results can be found in Appendix F.

### 6.2 Micro benchmarks

**Branch divergence.** Figure 3 shows the throughput of transaction execution with and without grouping on the transaction types. The bulks are generated for advance, and transactions are executed in parallel. We vary the number of transaction types (i.e., the number of branches in the *switch* clause). We denote "_L" and "_H" to be transac-

319

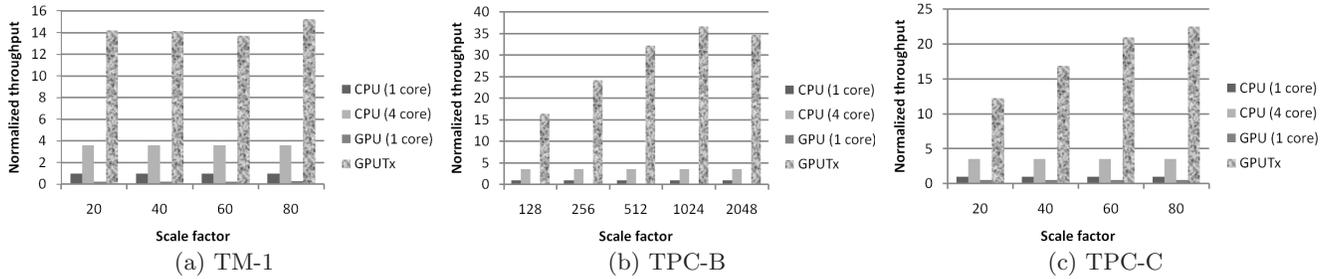

Figure 7: Normalized throughput for the public benchmarks.

tions with low and high computational costs ($x = 1$ and $x = 16$, respectively). Note that both x- and y-axis are given in the log scale. The measurement for the grouping method is obtained with the tuning on the number of radix partitioning passes. By grouping the transactions according to their branch, branch divergence in the SPMD execution is minimized. Thus, we observe that grouping reduces the branch divergence in the SPMD execution model of the GPU and improves the throughput, especially when the amount of computation is high and the number of branches is large.

We also find a cross point between the basic execution and the grouping method in the low-cost transactions, and there is no cross point in the high-cost transactions. For the low-cost transactions, when the number of branches is small, the impact of the sequential execution resulted from the divergent branches is small. The grouping overhead (typically with one pass of radix partitioning) further offsets this relatively small performance gain. In contrast, for the high-cost transactions, the grouping method is a clear winner even when the number of branches is two.

**Execution strategies.** Figure 4 shows the throughput of the three execution strategies as the bulk size varied. The number of tuples is fixed to be eight millions. As the number of transactions increases, the throughput of TPL decreases due to the increased contention of locks. In contrast, PART and K-SET achieve a stable and comparable throughput. The gap between TPL and the other two execution strategies increases. K-SET is slightly faster than PART, because PART has a larger runtime overhead (e.g., reading the boundary of partitions).

Figure 5 shows the time breakdown of executing around 16 million transactions. The total elapsed time is divided into two parts, *sort* for bulk generation, and *execution* for bulk execution. The bulk generation is a significant part for PART and K-SET, contributing 66% and 70% of the total elapsed time. In contrast, the transaction execution is a bottleneck for TPL, which contributes to 70% of the total execution time.

Figure 6 shows the throughput of the three execution strategies varying the transaction skewedness. TPL and PART naively pick the transactions in the transaction pool as a bulk, which generates a deep T-dependency graph. By extracting the 0-set continuously from the transactions in transaction pool, K-SET achieves a stable throughput. This significant improvement in throughput of K-SET has two reasons. First, the 0-set is sufficiently large for K-SET execution, as new transactions are submitted. Second, the GPU utilization of TPL and PART is lower than that of K-SET, due to the long critical path in the bulk execution.

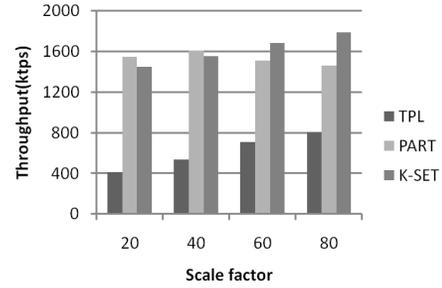

Figure 8: Throughput of the three execution strategies on TM-1.

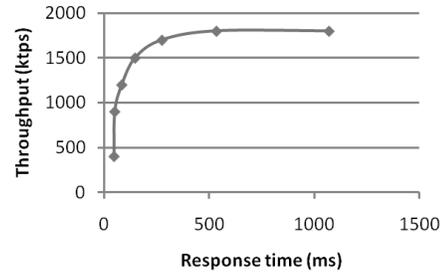

Figure 9: Response time vs. throughput in TM-1.

## 6.3 Public benchmarks

We simulate the ad-hoc transaction executions on the GPU by evaluating the transaction sequentially using one GPU core. The CPU-based counterpart adopts the design of H-Store [18]. Figure 7 shows the normalized throughput for GPUTx. All the measured throughput are normalized to the CPU-based engine on the single core. GPUTx on a single GPU core achieves a throughput of 25–50% as that of the CPU-based counterpart on a single CPU core. This is because, a single CPU core has a higher frequency as well as a higher memory bandwidth than a single GPU core. Devoting the major die size to cores, the GPU has much more cores than the CPU. While a single core has a lower throughput, the GPUTx has a much higher throughput than its CPU-based counterparts. Specifically, GPUTx achieves 4–10 times higher throughput than the CPU-based counterpart. Moreover, the improvement increases as the scale factor increases in the benchmark.

We further study the cost efficiency of GPUTx and its CPU-based counterpart. The unit prices of the NVIDIA GPU and the Intel CPU used in this experiment are US$1,699.00 and US$ 649.00, respectively[1]. We define the cost efficiency to be the throughput per dollar. While the GPU is more expensive than the CPU, GPUTx achieves 52%, 214%, and 98% on average higher throughput per dollar than the CPU-based implementation on TM-1, TPC-B and TP-C, respectively. Moreover, the GPU is an integral component in modern machine, and utilizing the GPU is able to increase the cost efficiency of the entire machine.

---
[1]http://www.dell.com/, Nov-15, 2010

We study some detailed results for GPUTx with TM-1. We have obtained similar results on the three benchmarks.

Figure 8 shows throughput of the three execution strategies on TM-1. As the scale factor increases, the 0-set becomes large, and K-SET achieves a higher throughput than the other two methods. TPL underperforms the other two methods for all the tested scale factors.

Figure 9 shows the average response time and the throughput of GPUTx on TM-1 with the scale factor 80. We consider the scenario of 1 million transactions per second to assess the capability of GPUTx. Transactions are submitted to the system uniformly in time. After a fixed time interval $t$, we generate a bulk from the transaction pool for execution. As the time interval increases, both the average response time and the throughput increase. The throughput increases sharply at the beginning. If the application can tolerate an average response latency of 534 ms, GPUTx achieves the best throughput on TM-1.

## 7. LIMITATIONS

Current GPUTx has many limitations, including:

**Support for pre-defined stored procedures only:** Current GPUTx does not allow users to compose the transactions command by command. It is suitable for applications with transactions, which can be described as a relatively static set of stored procedures.

**T-dependency graph construction:** The efficiency of bulk execution strategies in GPUTx depend on the parallelism exposed in T-dependency graph. The current method of constructing the T-dependency graph exploits the derivation on the affected data items. If transactions become more complicated [3], more advanced algorithms should be developed for the applicability of GPUTx.

**Sequential workload:** If the transaction workload is inherently sequential, e.g., the database with a single partition, only a limited number of cores on the GPU are utilized. As a single core on the GPU is slower than that on the CPU, GPUTx underperforms its CPU-based counterpart. A coprocessing scheme may be helpful to leverage the advantages of both the CPU and the GPU.

**Database fitting into the GPU memory:** Current GPUTx works correctly on the database fitting into the device memory. If the database cannot fit into the GPU memory, the data need to be copied on demand.

**GPUs without atomic operation support:** TPL requires GPUs with atomic operation support. Most desktop and laptop GPUs manufactured after 2009 have these capabilities. Nevertheless, PART and K-SET can work correctly for the GPUs without atomic operation.

**Portability to other many-core architectures:** Our current implementation is based on CUDA. It is our future work to evaluate our design and implementation on other many-core architectures such as AMD GPUs and Fusion.

## 8. CONCLUSIONS

The productivity of OLTP systems becomes an important performance factor for traditional and emerging online services. High-throughput transaction processing techniques are definitely beneficial to the user experiences and productivity in those applications. In this paper, we propose, GPUTx, a high-throughput transaction execution engine on the GPU. The design and implementation of GPUTx targets at optimizations of the massive parallelism of the GPU. We study three basic execution strategies for the bulk execution, with and without locks. Experimental results show that GPUTx achieves a 4–10 times higher throughput than its CPU-based counterpart on a quad-core CPU. As for future work, we are addressing the limitations of GPUTx for a complete transaction processing engine on the GPU.

## Acknowledgements

The authors thank the anonymous reviewers for their insightful suggestions. This work was supported by an AcRF Tier 1 grant from Singapore, an NVIDIA Academic Partnership (2010–2011), and a grant No. 419008 from the Hong Kong Research Grants Council.

# APPENDIX

## A. PRELIMINARY ON GPUS

GPUs, originally designed for graphics rendering tasks, have evolved into massively multi-threaded many-core co-processors for general-purpose computing. The GPU consists of many SIMD (Single Instruction, Multiple Data) multiprocessors, all sharing a piece of device memory. The execution on multi-processors is further organized in the SPMD execution model.

NVIDIA CUDA, a popular GPU programming framework, exposes the hierarchy of GPU threads. Warps, each of which consists of the same number of threads, are basic scheduling units across multiprocessors. Within a multiprocessor, warps are further grouped into thread blocks. CUDA also exposes the memory hierarchy. Multiprocessors share the device memory, which has a high bandwidth and a high access latency.

Branch divergence is an important performance factor. If threads of a warp diverge via a conditional branch, the warp sequentially executes each branch path taken. When all paths complete, the threads converge back to the same execution path. In the SPMD model, branch divergence occurs only within a warp; different warps execute independently regardless of their code paths.

## B. T-DEPENDENCY GRAPH CONSTRUCTION

The T-dependency graph is constructed by adding transactions one by one in the increasing order of their timestamps. Upon adding a new transaction, we add a vertex to the T-dependency graph and edges to the vertex. We denote the new transaction as $t$ and the new created vertex as $n$. Since $t$'s timestamp is larger than the ones in the T-dependency graph, we need to add an edge from a vertex in the T-dependency graph to $n$, if they satisfy the three requirements.

To facilitate the process of identifying the conflicting transactions, we introduce a data-oriented approach to speedup the search. The basic idea is to examine the transactions only when they access the same data item. We maintain a map to associate a data item with a list of transaction IDs in the ascending order, representing the list of transactions accessing the data item. For adding a new transaction, we need to check the transaction lists associated with all its basic operations. For the target data item of each basic operation, if the transaction list is empty, we simply add the transaction ID to the list and no edge is added. Otherwise, the checking process depends on whether the basic operation is a read or a write. If it is a write, we scan from the tail of the list until we meet a transaction $t_w$ with a write on the data item. If $t_w$ is the tail, we simply add an edge $t_w \rightarrow t$. Otherwise, we add an edge $t' \rightarrow t$, where $t'$ is a transaction (with a read on the data item) between the tail and $t_w$ (exclusive). If the basic operation is a read, we simply need to add an edge $t_w \rightarrow t$, no matter whether $t_w$ is tail or not.

The T-dependency graph construction requires the knowledge of the affected data item. This is particularly suitable for transactions with primary key accesses and for applications with tree-shaped schema, e.g., transactions in the public benchmarks. We consider transaction rewrites such as split [3] to transform the transactions suitable for T-dependency graph construction whenever practical. In the worst case, if the transaction conflicting relationship cannot be determined on the data item level, we determine the conflict at a coarser granularity, e.g., column or table.

```
bool leaveLoop=false;
while(!leaveLoop)
{
  int lockValue=atomicCAS(lockAddr, 0, 1);
  if(lockValue==0)
  {
        leaveLoop=true;
        //processing on the shared data
        *lockAddr=0;
  }
  __threadfence();
}
```

**Figure 10: The kernel code for a basic 0/1 spin-lock.**

```
/*initialize the values of all the lock to be zero. */
/*a GPU thread holds a <keyValue> for a tuple. */
bool leaveLoop=false;
while(!leaveLoop)
{
  volatile int lockValue=*lockAddr;
  if(lockValue==keyValue)
  {
        leaveLoop=true;
        //processing on the shared data
        if(flag==marked)
                atomicAdd(lockAddr, 1);
  }
  __threadfence();
}
```

**Figure 11: The kernel code for a counter-based lock.**

## C. IMPLEMENTATION OF SPIN-LOCK ON THE GPU

Figure 10 shows the source of the kernel code for a basic spin-lock implemented with atomic operations in CUDA. We use two APIs in CUDA for synchronization: (a) *atomicCAS(addr, compare, val)* is a compare-and-swap operation. It reads *addr* (let the value be *old*), computes (*old == compare ? val : old*), and stores the result back to memory at the same address. The function returns the *old* value. (b) *__threadfence()* is a barrier to ensure the data updates in the memory are visible to all the threads on the GPU. The problem of the basic lock is that it may have deadlock and non-deterministic execution.

Figure 11 shows the source of the kernel code for the spin-lock with deterministic execution. We use one API and one variable type qualifier for the implementation: (a) *atomicAdd(lockAddr, val)* increases the value in *lockAddr* by *value*; (b) *volatile* indicates the compiler that the variable value can be changed at any time by another thread and therefore any reference to this variable compiles to an actual memory read instruction.

## D. OPTIMIZATION ISSUES IN GPUTX

**Branch divergence.** Transactions of different types take different branches in the combined kernel, causing branch divergence on the GPU execution. Since the SPMD execution model limits the branch divergence within a warp, we can group the transactions in the bulk according to their trans-action type such that the branch divergence does not occur within the group. A naive grouping is to perform a radix sort on the transaction type ID. For $T$ transaction types, the bulk after sorting generates $T$ groups of transactions with the same type. Radix sort is a multi-pass algorithm, where each pass uses $b$ bits for sorting. The naive grouping method takes $\frac{\log_2 T}{b}$ passes of radix partitioning.

The naive algorithm is not designed to be aware of the different performance gains of each pass. As more passes are performed on the bulk, the performance gain on branch divergence reduction diminishes. We can stop the radix partitioning earlier for the optimal performance. Since all the transaction types are known in the system, we run calibration to determine the number of passes of radix partitioning. This calibration is sufficient, demonstrating performance improvement on the naive algorithm.

**Choosing the suitable execution strategy.** These three execution strategies have their own strength and weakness. TPL is a general method, with a relatively high runtime overhead, and the bulk size generated for TPL depends on the transaction coming rate. In contrast, K-SET has little runtime overhead, and the bulk size depends on the 0-set of the T-dependency graph. PART is in the middle, but the efficiency is subject to the single-partition requirement. All the three methods require sort operations, and TPL and K-SET has a relatively high bulk generation cost.

In particular, we identify the following three structural parameters of the T-dependency graph, as important indicators for the performance of bulk execution:

1. The depth of the T-dependency graph, $d$;
2. The number of transactions in 0-set, $w_0$;
3. The number of vertexes that have more than one preceding vertex (e.g., cross-partition transactions), $c$.

The $w_0$ value is an important indicator for the parallelism. Since each core on the GPU can execute one transaction at a time, executing a $k$-set of smaller than $M$ transactions is likely to underutilize the GPU computation resource ($M$ is the number of processors on the GPU). The depth of the T-dependency graph is the length of the critical path of the bulk execution, which is used as a good indicator on the elapsed time.

We develop a rule-based method to decide the suitable execution strategy, as described in Algorithm 1. If the number of transactions in the 0-set is so large that their executions can fully utilize the GPU resource (the threshold denoted as $\bar{w}_0$), K-SET is preferable, since it is with little overhead in run time and the load tends to be more balanced through fine-grained threading on the GPU. On the other hand, if the number of transactions in the 0-set is very small, we consider TPL and PART. The choice between TPL and PART depends on the number of cross-partition transactions and the depth. If the depth is larger than $\bar{d}$ or the number of cross-partition transactions is smaller than $\bar{c}$, we choose PART. Otherwise, TPL is chosen.

**Logging.** Logging is an important performance factor in transaction execution. Clearly, logging should be eliminated whenever practical. Re-do logging is the key for durability. Durability is not our focus in this study, and applications may achieve durability with non-logging methods, such as replications on multiple machines.

As for un-do logging, we plan to distinguish the transaction types with and without requirement on un-do log-

**Algorithm 1** Choosing the suitable execution strategy

1: Obtain the number of transactions in 0-set (let it be $w_0$);
2: **if** $w_0 \geq \bar{w}_0$ **then**
3:    Return K-SET;
4: **else**
5:    Let $c$ be the total number of cross-partition transactions in the bulk;
6:    Let $d$ be the depth of the T-dependency graph of the bulk;
7:    **if** $c \leq \bar{c}$ or $d \geq \bar{d}$ **then**
8:      Return PART;
9:    **else**
10:      Return TPL;

ging. Similar to H-Store [18], we write *two-phase* transactions when applicable for eliminating the undo log. In the first phase, it contains read-only operations. The transaction may be aborted, based on the result of these operations (e.g., record does not exist). In the second phase, the transaction performs a collection of operations without abortion. Since it performs abortion before updating the database, no un-do logging is required.

If there exists a transaction type that is not two-phase, we identify the set of transaction types that can be conflicting with it. We write un-do logs for the transactions belonging to those types. The log is written to in the GPU memory, and is discarded when the transactions commit.

When the transaction aborts, GPUTx runs the log-based recovery. The recovery process rolls back the updates from the aborted transactions as well as from the transactions in the sub-DAG of the T-dependency graph rooted at the transaction. For PART and K-SET, the recovery affects the transaction itself, since the conflicting transactions have not been executed in PART, or there are no conflicting transactions in the bulk of K-SET. For TPL, data operations from some conflicting transactions can be executed concurrently. We mark the transaction that requires recovery. After the bulk execution completes, the system rolls back the updates sequentially for those marked transactions. This recovery process is very costly for TPL.

## E. DETAILS ON EXPERIMENT SETUP

The CPU has 8MB shared L3 cache and four cores, each running at the frequency 2.26 GHz. Each GPU has 240 cores, each running at 1.3 GHz, and 4GB DDR3 device memory at 800 MHz. The measured peak memory bandwidth of device memory is 73GB/sec. The GPU is connected to the host machine with a PCI express, with a measured peak bandwidth of 3.4 GB/sec. We use NVIDIA CUDA v3.1 to implement GPUTx.

The database are initially loaded into the GPU memory, and the initialization cost is excluded for the throughput measurement. The throughput measurement includes the data transfer between the device memory and the main memory for the input transaction signatures and result output.

**Public benchmarks.** TM-1 is a telecom workload benchmark originally developed by Nokia. It consists of seven predefined transactions that insert, update, delete and query tuples from four large tables in the database. The subscriber ID is used as the partitioning key. In transactions UPDATE_LOCATION, INSERT_CALL_FORWARDING, and DELETE_CALL_FORWARDING, the subscriber is accessed with the string representation of the subscriber ID. The mapping from the string representation and the subscriber ID is static. We split each of these transactions into two: one searching the subscriber ID using the string representation, and the other for the remainder logic in the original transaction.

TPC-B is a database stress test consisting of a single transaction type. The branch ID is used as the partitioning key.

TPC-C approximates the workloads in an online transaction processing database for a retailer. It consists of five kinds of transactions, which model the process of customer orders from the initial creation to the final delivery and payment. The combined key of the warehouse ID and the district ID is used as the partitioning key. The payment and ostat transactions may search the customer using the last name. We split these two transactions into two: one searching the customer ID using the last name, and the other for the remainder logic in the original transaction. Fekete et al. [3] have provided a complete static analysis on the conflicts between transactions in TPC-C. We adopt their analysis in the T-dependency graph construction.

These benchmarks represent different characteristics to assess the optimization opportunities of GPUTx. They have different numbers of transaction types for assessing branch divergence. Moreover, transactions in TPC-B and TM-1 are short, usually accessing only 1-4 database rows, whereas those in TPC-C can access dozens of tuples. Finally, these benchmarks have different abortion rates. TM-1 has a higher abortion ratio than TPC-B and TPC-C.

Given the scale factor $f$ in the public benchmark, the maximum number of partitions in PART is $f$ million, $f$, and $f \times 10$ for TM-1, TPC-B and TPC-C, respectively. The actual number of partitions used for bulk execution is tuned according to the partition size. The suitable partition size is 128 in our experiment. We use ($f$ million/128), $f$, and $f \times 10$ for TM-1, TPC-B and TPC-C, respectively.

**Implementation.** We apply column-based stores as the storage on the GPU, in order to minimize the data transfer between the main memory and the device memory. Our current column-based store adopts the simple format. If the column is fixed-length, GPUTx stores the values of the column in an array, and the column-wise information such as the length of the column are stored in the system catalog. Otherwise, GPUTx represents a tuple as the offset and the length, where the offset points to the start position of the tuple in an array storing the values. The column-based storage allows copying only the required columns to the GPU. Another benefit is that the GPU can support a larger scale of the transaction workloads than that of the row-based store.

When GPUTx is up, the database data is loaded into the main memory. Next, the necessary data columns and indexes are copied from the main memory to the GPU memory, and GPUTx is ready to accept transactions. After a bulk is generated, its parameter values are copied to the GPU memory for the bulk execution. The result construction (usually projections) after the bulk execution is performed on the main memory. The results of the bulk execution include the result record IDs and the values of the columns that are not in the main memory. Based on these results, we construct the results for the bulk based on the columns stored in the main memory. Read-only columns are stored in the main memory to facilitate result constructions.

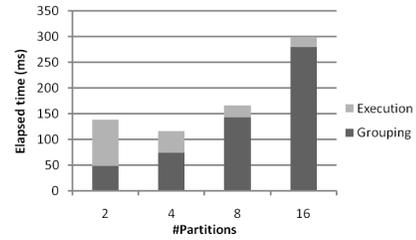

Figure 12: Time breakdown for execution.

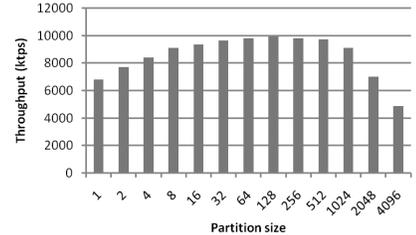

Figure 13: Throughput varying the partition size in PART.

We attempt to reduce the amount of efforts for implementing a specific application on GPUTx. Components like $k$-set determination and the bulk execution strategies are general-purpose ones for all applications. There are some components in GPUTx required the expertise from DBA: (a) annotations on whether the transaction is single-partition or not, and the partitioning key; (b) transaction rewrites [3] to facilitate the T-dependency graph construction; (c) domain-specific rules on detecting whether two transactions are conflicting.

## F. MORE EXPERIMENTAL RESULTS

### F.1 More Results on Micro Benchmarks

**Grouping on branch divergence.** Figure 12 shows the time breakdown of the grouping-based method ($x = 32$ and $T = 16$) with varying the number of partitions. We divide the elapsed time into two parts, *grouping* and *execution* on transactions. As the number of partitions approaches to the number of branches (i.e., more radix partitioning passes), the execution time significantly reduces due to the reduction on the branch divergence, but the grouping overhead increases as well. The optimal $p$ value is four for this setting.

**Tuning the partition size in PART.** Figure 13 shows the throughput of PART transactions with $x = 16$. Confirming our cost analysis, the curve of the throughput is a concave one, as the partition size increases. The optimal partition size is 128 in this experiment, as a combined effect from the amount of computation of each thread, the thread parallelism as well as the overhead of the auxiliary structure in PART.

**Varying the number of tuples.** Figure 14 shows the throughput varying the number of tuples. The number of transactions in the transaction pool is fixed to be 256 thousands. As the number of tuples becomes large, the probability of conflicting transactions becomes small. Therefore, as the number of tuples increases, all the three execution strategies achieve a higher throughput. However, the reasons for the improved throughput are different. The improvement of TPL is mainly due to the reduction in the lock contention.

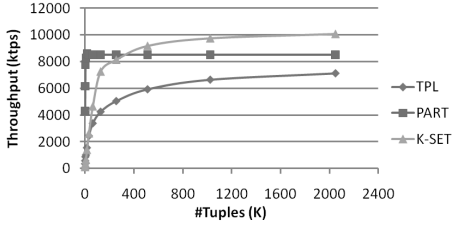

Figure 14: Throughput of the three transaction execution strategies varying the relation cardinality.

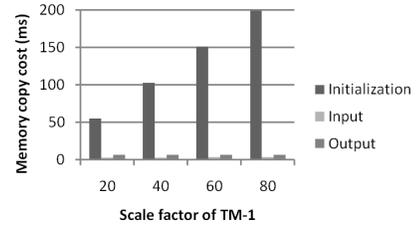

Figure 16: Memory transfer cost between the GPU memory and the main memory on TM-1.

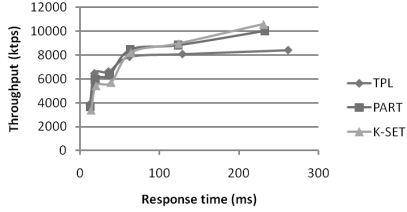

Figure 15: Response time vs. throughput in micro benchmarks.

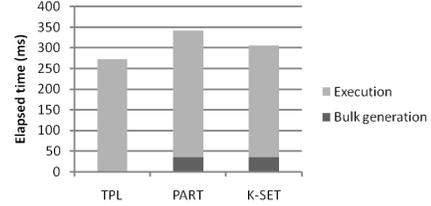

Figure 17: The time breakdown without timestamp constraint for micro benchmarks.

The improvement of PART is mainly due to the shorter critical path of the bulk execution. As the number of tuples increases, K-SET extracts a larger 0-set to achieve a better utilization of the GPU resource.

**Response time vs. throughput.** Figure 15 shows the average response time and the throughput of three execution strategies. We consider the scenario of 4 million transactions per second and transactions are submitted to GPUTx uniformly in time. The throughput of the three strategies reaches the peak when the average response time is larger than 260 ms. The bulk size is small when the the interval is small, which results in a small 0-set. In this case, TPL is the best. As the interval increases, PART and K-SET achieves a higher throughput than TPL.

### F.2 More Results on Public Benchmarks

**Column- vs. Row-based storage.** The column-based storage has its advantages over the row-based storage in GPUTx. First, in TM-1, by storing only the necessary columns on the GPU, the column-based storage reduces the amount of GPU memory consumption by 27% for different scale factors. In specific, when the scale factor is 80, the tables and indexes in the row- and the column-based storages consume 1, 756 MB and 1, 279 MB, respectively.

Second, GPUTx with the column-based storage is around 10% faster than that with the row-based storage. The major reason is that the column-based storage increases the locality of memory accesses among different threads. Due to the SPMD execution, a column tends to be accessed by many threads at the same time. Basic memory operations like gather and scatter have a higher memory performance on the fine-grained column-based storage than those on the row-based storage.

**Memory transfer.** Figure 16 shows the data transfer cost between the GPU memory and the main memory of running TM-1 on GPUTx. There are three cost components: (1) "initialization": during initialization of GPUTx, database data (tables and indexes) are loaded into the GPU memory; (2) "input": before a bulk execution, the bulk parameter values are copied from the main memory to the GPU; (3) "output": after a bulk execution, the results are copied from the GPU memory to the main memory for further result construction. The initialization is performed *once* when the system is up. The latter two components are included in the bulk execution, and they contribute to less than 5% of the total execution time.

## G. RELAXING TIMESTAMP CONSTRAINT

The correctness definition of bulk execution poses a timestamp constraint on the bulk execution. This timestamp constraint results in a sort operation on the bulk generation, and also restricts the parallelism of the bulk execution. Some applications may require the basic serializability without the timestamp constraint. Relaxing the timestamp constraint offers the opportunities in reducing the runtime overhead of bulk generation and execution. We briefly discuss our preliminary efforts in revisiting the three basic execution strategies without timestamp constraint.

Without the timestamp restriction, TPL uses the basic spin lock (as shown in Figure 10). The dependency among transactions is captured with the classic dependency graph.

The bulk generation of PART and K-SET can also be simplified. Similar counter based methods are developed for PART and K-SET, and we focus on PART. The PART method is to assign a lock on each partition. Each lock has a counter value (initially zero). During the bulk generation, each transaction needs to acquire the lock for its partition, get the counter value as its key value, and increases the counter value by one. The final counter value of each spin lock is the number of transactions for the corresponding partition. A prefix sum is used to calculate the start position of each group in an array. The key value of each transaction is the relative position in the partition. Thus, transactions can be grouped without sort.

Figure 17 shows the time breakdown of the three execution strategies without timestamp constraint for micro benchmarks. The experimental setup is the same as that of Figure 5. The cost for both bulk generation and execution significantly reduces. In this experiment, the locking overhead is small, and TPL outperforms the other two execution strategies, which is in contrast with Figure 5.